\documentclass[sigconf]{acmart}
\AtBeginDocument{%
  }
\usepackage{listings}
\usepackage{comment}
\usepackage{xcolor}
\usepackage[most]{tcolorbox}
\usepackage{booktabs}
\usepackage{needspace}
\usepackage{placeins}
\tcbuselibrary{skins,breakable,listings}
\usepackage{enumitem}
\usepackage{hyperref}
\tcbuselibrary{breakable,listings}
\usepackage{siunitx}
\usepackage{graphicx}
\usepackage{refcount}
\lstdefinestyle{jsonstyle}{
  basicstyle=\ttfamily\small,
  breaklines=true,
  frame=single,
  numbers=none,
  showstringspaces=false,
  tabsize=2,
  columns=fullflexible
}
\lstdefinestyle{pystyle}{
  language=Python,
  basicstyle=\ttfamily\footnotesize,
  breaklines=true,
  frame=single,
  numbers=none,
  showstringspaces=false,
  tabsize=2,
  columns=fullflexible
}
\acmConference[EASE '26]{International Conference on Evaluation and Assessment in Software Engineering}{June 09--12, 2026}{Glasgow, UK}

\begin{document}

\title{Understanding the Limits of Automated Evaluation for Code Review Bots in Practice}

\author{Veli Karakaya}
\authornote{Both authors contributed equally to this research.}
\email{veli.karakaya@ug.bilkent.edu.tr}
\author{Utku Boran Torun}
\authornotemark[1]
\email{boran.torun@bilkent.edu.tr}
\affiliation{%
  \institution{Bilkent University}
  \city{Ankara}
  \country{Turkey}
}

\author{Baykal Mehmet Uçar}
\email{baykal.ucar@beko.com}
\affiliation{%
  \institution{Beko}
  \city{İstanbul}
  \country{Turkey}
}

\author{Eray Tüzün}
\email{eraytuzun@cs.bilkent.edu.tr}
\affiliation{%
  \institution{Bilkent University}
  \city{Ankara}
  \country{Turkey}
}

\settopmatter{printacmref=true}
\begin{abstract}
Automated code review (ACR) bots are increasingly used in industrial software development to assist developers during pull request (PR) review. As adoption grows, a key challenge is how to evaluate the usefulness of bot-generated comments reliably and at scale. In practice, such evaluation often relies on developer actions and annotations that are shaped by contextual and organizational factors, complicating their use as objective ground truth. We examine the feasibility and limitations of automating the evaluation of LLM-powered ACR bots in an industrial setting. 
We analyze an industrial dataset from Beko comprising 2,604 bot-generated PR comments, each labeled by software engineers as fixed/wontFix. Two automated evaluation approaches, G-Eval and an LLM-as-a-Judge pipeline, are applied using both binary decisions and a 0–4 Likert-scale formulation, enabling a controlled comparison against developer-provided labels. Across Gemini-2.5-pro, GPT-4.1-mini, and GPT-5.2, both evaluation strategies achieve only moderate alignment with human labels. Agreement ratios range from approximately 0.44 to 0.62, with noticeable variation across models and between binary and Likert-scale formulations, indicating sensitivity to both model choice and evaluation design. Our findings highlight practical limitations in fully automating the evaluation of ACR bot comments in industrial contexts. Developer actions such as resolving or ignoring comments reflect not only comment quality, but also contextual constraints, prioritization decisions, and workflow dynamics that are difficult to capture through static artifacts. Insights from a follow-up interview with a software engineering director further corroborate that developer labeling behavior is strongly influenced by workflow pressures and organizational constraints, reinforcing the challenges of treating such signals as objective ground truth.
\end{abstract}


\maketitle

\section{Introduction}

Code review plays a central role in modern software development, serving as a critical quality assurance mechanism for detecting bugs, improving maintainability, and transferring knowledge among developers \cite{understandingACRprocessandDeveloper, Badampudi_2019, yang2024survey}. To reduce manual review overhead and accelerate development workflows, many organizations have adopted Automated code review (ACR) bots such as ReviewBot \cite{cihan2024automated}, SonarLint \cite{sonarqube_ide_2026}, and BitsAI-CR \cite{bitsaicr2025}, which generate comments on pull requests (PRs) to highlight potential issues or improvements. These bots are designed to augment human reviewers, yet their practical usefulness and reliability remain uncertain, particularly because organizations lack reliable ways to evaluate the value of bot-generated feedback in industrial settings.

While ACR bots can provide rapid and scalable feedback on PRs \cite{WesselAdoptCodeReviewBot}, their practical value critically depends on how the quality of the generated review comments is assessed in practice. Prior work by Cihan et al. \cite{cihan2024automated} in industrial settings shows that a 33.4\% of bot comments can be incorrect, misleading, or not applicable, highlighting the challenges of using developer actions as a reliable proxy for comment usefulness. Developer responses to bot comments often reflect not only technical validity, but also prioritization pressures, workflow constraints, and timing considerations. Low-quality or poorly timed feedback can waste developer time, introduce unnecessary changes, and ultimately reduce trust in review automation. As a result, organizations must solve not only how to deploy ACR bots, but also how to determine whether their feedback genuinely helps developers during PR review  \cite{understandingACRprocessandDeveloper}. This observation is further supported by our interview with an director of software engineering at Beko, who noted that developers’ responses to bot comments are frequently shaped by workflow pressures and prioritization constraints rather than purely technical considerations.

\begin{figure}[h]
    \centering
    \includegraphics[width=0.8\linewidth]{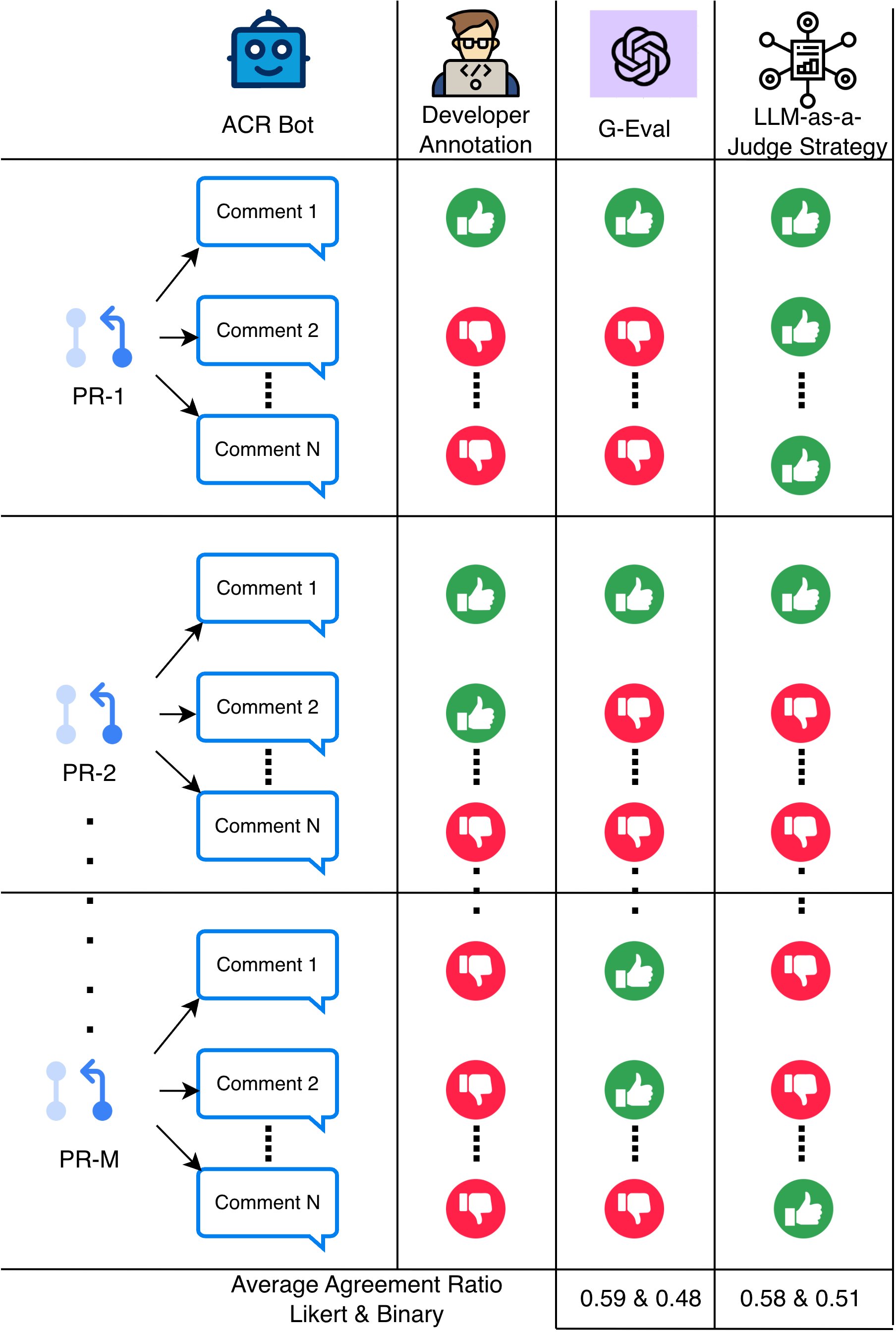}
    \caption{This figure illustrates how multiple evaluation strategies assess the usefulness of ACR bot comments across PRs (PR-1...PR-M) against the human labeled outcomes.}
    \label{fig:evaluation_strategies}
\end{figure}

Despite the widespread adoption of ACR bots, reliably evaluating the quality and usefulness of their generated review comments remains an open problem in industrial practice \cite{cihan2024automated}. Commercial ACR systems continuously produce large volumes of feedback, yet there is no standardized, scalable, and reliable mechanism to determine whether these comments meaningfully assist developers or merely introduce review noise \cite{khan2026surveycodereviewbenchmarks}. In current practice, effectiveness is primarily assessed through human judgment, such as manual inspection or developer feedback \cite{understandingACRprocessandDeveloper}. However, human-centered evaluation remains expensive, slow, and difficult to repeat at scale \cite{torun2026vision}. This becomes especially problematic as ACR systems evolve through model updates, prompt changes, or rule modifications \cite{cihan2024automated,bitsaicr2025}. In addition, developer actions such as fixing or ignoring comments are imperfect labels because they may reflect local priorities, release pressure, ownership boundaries, or timing rather than usefulness alone \cite{cihan2024automated}. These challenges are further amplified by the increasing adoption of commercial ACR solutions such as CodeRabbit \cite{coderabbit_ai}, Qodo \cite{qodo_ai}, and Snyk Code \cite{snyk_code_docs}, which makes systematic comparison across tools and versions particularly difficult \cite{google-ai-assisted-modern-code-review}. Figure \ref{fig:evaluation_strategies} illustrates the automated evaluation setting considered in this study, where multiple evaluation strategies independently assess bot-generated comments and are compared based on their agreement with human-labeled annotations.


Evaluating ACR comments is fundamentally challenging because usefulness is inherently subjective and depends heavily on the pull request context, including the code diff, developer intent, and project conventions. As a result, prior studies often rely on manual inspection, expert judgment, or partially automated procedures \cite{bitsaicr2025, cihan2024automated, chatGPTinautomatedcoderefinement, fine-tuning-llmsfor-codereview-automation}. Although informative, these approaches are time-consuming, inconsistent across evaluators, and difficult to scale to large industrial repositories \cite{cihan2024automated}. This creates a practical gap: industry requires evaluation methods that are not only accurate, but also scalable, consistent, and cost-effective. Whether such evaluation can be fully automated without oversimplifying developer judgment and contextual nuance remains an open question.

This challenge is closely related to the broader problem of evaluating LLM outputs. As recent ACR systems increasingly leverage LLMs to generate review feedback \cite{chatGPTinautomatedcoderefinement, fine-tuning-llmsfor-codereview-automation}, assessing ACR usefulness naturally depends on advances in LLM evaluation methodologies. Recent work proposes rubric-based, reference-free evaluation protocols (e.g., G-Eval) \cite{liu-etal-2023-g} and investigates the reliability of LLM judges \cite{zhang-etal-2024-benchmarking, chen-etal-2024-humans}. Benchmarking and survey studies further highlight the diversity of evaluation designs \cite{gao-etal-2025-llm}, while retrieval-augmented evaluation frameworks such as RAGAS emphasize the role of grounding and faithfulness \cite{es-etal-2024-ragas}. Despite these advances, existing LLM evaluation frameworks are largely generic and are not specialized for software engineering tasks. Moreover, in industrial code review settings, human labels themselves may reflect workflow constraints, risk tolerance, and prioritization decisions, introducing additional ambiguity that automated evaluators are not designed to resolve.

To the best of our knowledge, no widely adopted task-specific automated evaluation framework currently exists for industrial ACR workflows. Moreover, it remains unclear whether such standardization is feasible in practice, given the inherently contextual, project-specific, and organizational nature of code review decisions.

This study examines the feasibility and limitations of automated evaluation strategies for judging the usefulness of bot-generated code review comments, with the goal of understanding where and why such automation breaks down in industrial settings, while explicitly acknowledging that certain design choices such as scoring weights and decision thresholds require human intervention. We conducted this study in collaboration with Beko, a multinational home appliances company, whose software teams have adopted an LLM-based automated code review bot in their industrial PR workflows. The dataset used in this study originates from prior work by Cihan et al.~\cite{cihan2024automated}, where a Qodo-based code review agent powered by GPT-4 Turbo \cite{openai_gpt4_turbo_model_docs_2026} was deployed in Beko’s industrial repositories. Using data collected from this real-world deployment, we investigate two automated evaluation strategies, G-Eval and an LLM-as-a-Judge pipeline, and analyze their agreement with developer-provided labels that are commonly treated as ground truth in practice. To contextualize and interpret our quantitative findings, we additionally conducted a semi-structured interview with the director of software engineering at Beko.

Our objective is to assess whether evaluation can be automated. This leads us to the following research questions.

\begin{tcolorbox}[
  enhanced,
  breakable,
  title=\textbf{Research Questions},
  colback=white,
  colframe=black!25,
  coltitle=black,
  fonttitle=\bfseries\small,
  boxrule=0.5pt,
  arc=1.2mm,
  left=1.2mm,right=1.2mm,top=0.8mm,bottom=0.8mm,
  toptitle=0.6mm,bottomtitle=0.6mm
]
\small
\textbf{RQ1:} To what extent does a rubric-based G-Eval strategy align with human-labeled annotations when assessing the usefulness of ACR bot comments?\\[1pt]
\textbf{RQ2:} To what extent does a plain LLM-as-a-Judge evaluation strategy align with human-labeled annotations when assessing the usefulness of ACR bot comments?
\end{tcolorbox}

Rather than assuming high alignment as a necessary or attainable objective, these research questions are designed to characterize the degree and nature of agreement between automated evaluators and developer-provided labels in an industrial setting.

\section{Related Work}\label{sec:section3}

\subsection{Automated Code Review}\label{sec:acr}
ACR research has progressed from early ML/DL models trained on historical PRs and review feedback to recent LLM-based systems that can generate and assess review comments with richer context. However, most of this literature emphasizes generation quality or deployment outcomes, while the problem of evaluating comment usefulness in practice remains underexplored. We therefore organize prior work around this gap.

\subsubsection{\textbf{Automated Code Review with ML and DL}}\label{sec:acr_ml}
Tufano et al.~\cite{tufano2021automating} used a neural machine translation (NMT) model trained on historical PRs to learn reviewer-style code transformations, evaluating it in both contributor-style edit prediction and comment-available refinement settings. Li et al.~\cite{li2022codereviewer} introduced \textit{CodeReviewer}, a transformer pre-trained on large multilingual code-change and review-comment data to support multiple review tasks (e.g., comment generation and code refinement), establishing a strong baseline across benchmark tasks. Li et al. also proposed \textit{AUGER}~\cite{li2022auger}, which generates review comments using pre-trained DL models and evaluates output quality with standard generation metrics and usefulness signals from human-grounded assessments. 

Hong et al.~\cite{hong2022commentfinder} presented \textit{CommentFinder}, a retrieval-based approach that recommends existing human review comments for similar code changes, emphasizing efficiency and practical deployability via fast candidate retrieval and ranking.

Alexander et al.~\cite{resolvingCodeReviewWithML} reported industrial-scale deployment of ML systems that automate the resolution of frequent review patterns to reduce manual reviewer effort and improve throughput, focusing on operational impact in real workflows.

\subsubsection{\textbf{Automated Code Review with LLMs}}\label{sec:acr_llm}
LLMs have enabled more context-aware ACR pipelines \cite{cihan2025evaluatinglargelanguagemodels, chatGPTinautomatedcoderefinement, fine-tuning-llmsfor-codereview-automation}, including hybrid systems, end-to-end assistants, fine-tuning methods, and retrieval-augmented prompting aimed at improving comment quality and applicability.

Sun et al.~\cite{bitsaicr2025} proposed BitsAI-CR, a two-stage hybrid pipeline that first detects issues with rule-based/static checks and then uses an LLM to validate and refine candidates, reporting precision and deployment-focused adoption metrics.

Guo et al.~\cite{chatGPTinautomatedcoderefinement} empirically studied ChatGPT for automated code refinement by comparing it against CodeReviewer on public and curated datasets, using generation/refinement metrics (e.g., EM, BLEU) and analyzing common failure modes tied to comment ambiguity and missing context.

Vijayvergiya et al.~\cite{google-ai-assisted-modern-code-review} introduced AutoCommenter, an end-to-end LLM-backed assistant that enforces coding best practices across multiple languages at Google, evaluating success primarily through large-scale deployment and developer-rated helpfulness.

Ramesh et al. \cite{ramesh2025automatedcodereviewusing} reports real-world experience with integrating LLM-based automated code review into industrial development workflows. Their experience report emphasizes deployment considerations and practical challenges that arise when moving from promising model outputs to reliable, everyday code review support.


Alexander et al.~\cite{resolvingCodeReviewWithML} also explored LLM-assisted review support and automated application of review feedback to code changes, emphasizing automation of fixes rather than explicit usefulness assessment of the comments.

Lu et al.~\cite{lu2023llamareviewer} proposed \textit{LLaMA-Reviewer}, adapting LLaMA for review comment generation via parameter-efficient fine-tuning (LoRA), and evaluated it on public code review datasets to measure generation performance under low trainable-parameter budgets.

Zhang et al.~\cite{zhang2025laura} proposed \textit{LAURA}, a retrieval-augmented prompting method that supplies LLMs with historical review context to improve generated comment quality, evaluating gains using correctness/helpfulness-oriented outcome measures.

These studies show how recent systems increasingly incorporate richer context and stronger models, but they still treat evaluation largely as a byproduct of generation quality.

\subsection{LLM Evaluation}\label{sec:llm_eval}
Recent work on LLM evaluation increasingly favors rubric-based and reference-free judging to better match human preferences in open-ended tasks, while also documenting systematic biases and reliability issues.

Liu et al.~\cite{liu-etal-2023-g} introduced G-Eval, which uses a strong LLM with rubric-guided prompting (including chain-of-thought style reasoning) to score generated text, aiming to align more closely with human judgments than surface-form overlap metrics.

Zheng et al.~\cite{zheng2023judging} studied the LLM-as-a-Judge paradigm at scale via MT-Bench and Chatbot Arena, examining agreement with human preferences and documenting evaluation risks such as bias and prompt sensitivity that affect judgment reliability.

Gao et al.~\cite{gao-etal-2025-llm} surveyed LLM evaluation methods and organized common protocols (e.g., rubric-based, reference-free, and pairwise comparison), providing a taxonomy that motivates judgment-oriented evaluation beyond exact-match metrics.





Wang et al. \cite{Wang_2025} empirically study whether LLM-as-a-Judge can replace human evaluators across software engineering tasks (e.g., code generation/translation/summarization) and show that alignment with human ratings is strongly task- and setup-dependent, limiting the applicability of a single judge configuration across SE settings.

He et al. \cite{he-et-al-LLM-as-a-Judge} present an SE-focused perspective on LLM evaluation by synthesizing evidence on LLM-as-a-Judge usage in software engineering and highlighting recurring reliability risks such as bias, context sensitivity, and calibration issues, motivating SE-specific evaluation designs.

Zhou et al. ~\cite{zhang-etal-2024-benchmarking} and Chen et al. ~\cite{chen-etal-2024-humans} analyzed judgment reliability and the relationship between automatic metrics and human ratings, reinforcing the need for evaluation setups that better reflect human perception. Together, these works motivate rubric-based, reference-free LLM evaluators for assessing the usefulness of LLM-generated code review comments against human judgments.

For code review evaluation, the most relevant studies are those that connect judge design to the ambiguity of review usefulness. Naik et al. \cite{naik2025crscoregroundingautomatedevaluation} proposed CRScore, a reference-free metric for evaluating code review comments along dimensions such as relevance and comprehensiveness, supporting the need for task-specific evaluation metrics for LLM-based code review systems. Cihan et al.~\cite{cihan2024automated}, in turn, are central for our problem framing because they show in an industrial setting that developer resolutions and comment outcomes are informative but imperfect signals of usefulness. 

Together, these studies motivate our central question: even with strong rubric-based judges, how far can automated evaluation go when the available ``ground truth'' is itself shaped by workflow context, prioritization, and organizational constraints?

\section{Methodology}\label{sec:section4}


\subsection{Data Source and Extraction}\label{sec:datasource-and-extraction}
We use an industrial dataset of 2,604 PR comments from Beko's repository previously analyzed by Cihan et al. \cite{cihan2024automated}. The dataset contains bot-generated PR comments and human-labeled outcomes (fixed / wontFix) derived from developer labeled comments. There were five labels (active, pending, resolved (fixed), wontFix, and closed) for each comment. We exclude comments labeled as active and pending because they represent intermediate workflow states where no final developer decision regarding usefulness has been made. We also exclude closed comments because, in the original dataset definition, this label indicates that a comment was not implemented for reasons other than wontFix (e.g., external or workflow-related factors) rather than as a result of an explicit technical judgment about usefulness \cite{cihan2024automated}. Since automated evaluators and LLM-based judges do not have access to these external decision factors, treating closed as not useful would introduce additional ambiguity into the ground truth. The resulting dataset consists of 1,733 comments ($\sim$66\%) labeled as fixed and 871 comments ($\sim$34\%) labeled as wontFix.

Data are accessed via the organization's Azure DevOps instance and Jira project. For each PR we extract: (1) PR Description, PR Title, PR diff (2) All bot-generated review comments (3) JIRA issue description if specified in PR.
Human-labeled usefulness is given and considered as ground truth of this study.

\paragraph{Human-Labeled Usefulness}
The usefulness labels used in this study originate from an existing industrial code review workflow. As reported by Cihan et al.~\cite{cihan2024automated}, developers are required to explicitly finalize the status of all review comments (e.g., fixed, wontFix, active, pending, closed) before a PR can be merged. This enforced process ensures that each review comment is deliberately assessed by developers during real code review activities.

Following this practice, we map comments marked as fixed to useful, and comments marked as wontFix to not useful. The definition of usefulness is given in replication package\footnote{\url{https://figshare.com/articles/conference_contribution/Understanding_the_Limits_of_Automated_Evaluation_for_Code_Review_Bots_in_Practice/31462948?file=62372143}}. These labels reflect developers’ real decisions during production code review and are therefore treated as ground truth for evaluating automated review comment usefulness.
\begin{figure}[t]
    \centering
    \includegraphics[width=1\linewidth]{figures/methodFigure.png}
    \caption{Evaluation setup comparing G-Eval and LLM-as-a-Judge (binary vs. 0–4 Likert) for labeling bot comment usefulness and reporting agreement and classification metrics.}
    \label{fig:methodFigure}
\end{figure}
\subsection{Evaluation Criteria}\label{sec:evaluation_criteria}

\textbf{Models.}
We evaluate both automated evaluators with three LLM backends: Gemini-2.5-pro \cite{google_gemini_2_5_pro_vertexai_2026}, GPT-4.1-mini \cite{openai_gpt_4_1_api_2025}, and GPT-5.2 \cite{openai_gpt_5_2_api_2025}.
To assess the usefulness of ACR comments, we evaluate each comment using two complementary rubric settings:
(i) 0-4 Likert scale and (ii) a direct binary usefulness rubric as shown in Figure \ref{fig:methodFigure}.
The Likert scale provides a structured breakdown of comment quality, while the binary rubric captures an overall usefulness decision aligned with the dataset’s useful/not useful labels.
\subsubsection{0-4 Likert Scale and Decision Rule.}\label{sec:Likertscale}
In the 0-4 Likert scale setting, the evaluator assigns integer scores
$s_{\text{usefulness}} \in \{0,1,2,3,4\}$ for ACR bot comments. The full definition of these usefulness scores are given in the replication package \cite{Karakaya2026}.

\paragraph{Rationale for the 0--4 Scoring Scale.}
We adopt a discrete 0--4 scoring scale to implement a simple and interpretable evaluation rubric. This design choice is consistent with prior LLM-as-a-Judge and rubric-driven evaluation frameworks, which emphasize explicit and well-specified scoring criteria to improve interpretability for both human annotators and language models~\cite{liu-etal-2023-g,chang2023surveyevaluationlargelanguage}. To enable comparison with binary developer labels, we map Likert scores to binary outcomes using a threshold of 1. In this mapping, a score of 0 denotes the absence of any perceived usefulness signal, while scores of 1 or higher indicate the presence of at least some signal, even if weak or imperfect. This interpretation is consistent with prior work showing that low non-zero Likert scores may still correspond to problematic or hallucinated content, yet nonetheless reflect the presence of a detectable response rather than a complete absence of information~\cite{tantithamthavorn2026hallujudge}.

\begin{tcolorbox}[
  breakable,
  enhanced,
  colback=gray!5!white,
  colframe=black!75!white,
  title={\textbf{Abbreviated rubric summary}},
  boxrule=0.5pt,
  arc=3pt,
  left=6pt, right=6pt, top=6pt, bottom=6pt
]
\textbf{0 (Not Useful):} Incorrect, misleading, or unrelated to the PR diff, \ldots{} not actionable.\\
\textbf{1 (Low Usefulness):} Minor/generic guidance with limited PR-specific grounding, \ldots{} limited impact.\\
\textbf{2 (Borderline Useful):} Plausibly relevant but evidence/actionability in this PR is unclear, \ldots{}.\\
\textbf{3 (Mostly Useful):} PR-specific non-obvious issue with enough evidence to act now, \ldots{}.\\
\textbf{4 (Fully Useful):} PR-critical and evidence-backed feedback that should be handled in this PR, \ldots{}.
\end{tcolorbox}

\subsubsection{Binary Usefulness Rubric (0--1)}\label{sec:binary_rubric}

In addition to the 0-4 Likert scale, we evaluate a direct binary usefulness decision to match the dataset’s ground truth framing.
In this setting, the evaluator assigns a single usefulness score constrained to the set $\{0,1\}$.

The binary prediction is obtained deterministically by mapping the usefulness score to the label: useful when the usefulness score is 1 and, not useful otherwise. The corresponding full explanations of scores are given in the replication package \cite{Karakaya2026}.
\begin{tcolorbox}[
  breakable,
  enhanced,
  colback=gray!5!white,
  colframe=black!75!white,
  title={\textbf{Abbreviated rubric summary}},
  boxrule=0.5pt,
  arc=3pt,
  left=6pt, right=6pt, top=6pt, bottom=6pt
]
\textbf{0 (Not Useful):} Incorrect/irrelevant/confusing feedback with no actionable value, \ldots{}.\\
\textbf{1 (Useful):} Technically correct, relevant, clear, and actionable feedback, \ldots{} improves code quality.
\end{tcolorbox}
\subsubsection{Shared Inputs, Prompting Scope, and Decision Mapping}\label{sec:shared_eval_setup}
Both evaluation strategies (G-Eval and LLM-as-a-Judge) use the same shared evaluation setup so that differences can be attributed to the evaluation approach itself.

\paragraph{Inputs.}
For each bot comment, the evaluator receives the same compact PR context: PR title, PR description, PR diff, and the bot comment. No repository-wide artifacts (e.g., project guidelines or historical discussions) are provided.

\paragraph{Prompting Scope.}
In both strategies, prompts instruct the evaluator to ground judgments in the provided PR artifacts and apply the same usefulness criteria defined in Sections~\ref{sec:Likertscale} and \ref{sec:binary_rubric}. The key difference from G-Eval is prompt style: LLM-as-a-Judge uses direct judge-style instructions rather than G-Eval's rubric-metric template but the underlying evaluation criteria remain identical.

\paragraph{Decision Mapping.}
For Likert-based evaluation, numeric scores are mapped to binary labels using the same thresholding rule in Section~\ref{sec:Likertscale} (useful iff $s_{\text{usefulness}} \ge 1$). For direct binary evaluation, outputs are deterministically mapped using Section~\ref{sec:binary_rubric} (useful iff score = 1).

\subsection{G-Eval Based Evaluation}\label{sec:gevalmethod}
Our first evaluation strategy is based on G-Eval\footnote{\label{fn:geval}\url{https://deepeval.com/docs/metrics-llm-evals}}, a rubric-driven LLM-based evaluation framework designed for reference-free and subjective assessment tasks. Evaluating the usefulness of ACR bot comments exhibits two key characteristics: (i) there is no single ``gold'' reference comment for a given PR diff, and (ii) usefulness is inherently context dependent, requiring reasoning over PR artifacts rather than surface-level lexical overlap. As a result, conventional reference-based automatic metrics (e.g., BLEU, ROUGE) are not applicable, and retrieval-focused evaluation frameworks such as RAGAS~\cite{es-etal-2024-ragas} do not directly capture the notion of comment usefulness.

G-Eval aligns with these requirements by operationalizing evaluation as an explicit rubric-following scoring procedure executed by an LLM. It enables structured scoring under well-specified criteria and can be applied without task-specific fine-tuning. Prior work reports that rubric-driven prompting with intermediate reasoning can improve alignment with human judgments in subjective evaluation settings~\cite{liu-etal-2023-g}. We therefore use G-Eval as a baseline to assess whether a rubric-based evaluator can recover developers' labels from PR artifacts alone.

This strategy uses the shared setup in Section~\ref{sec:shared_eval_setup} for inputs, criteria, and decision mapping.

\subsection{LLM-as-a-Judge Evaluation}\label{sec:llmasajudgemethod}

Our second evaluation strategy follows an LLM-as-a-Judge design, where an LLM is prompted to act as an evaluator and decide whether a bot-generated review comment is useful. This approach is motivated by the growing use of LLMs as meta-evaluators for subjective tasks, in which explicit references are unavailable and the assessment criteria are inherently context dependent \cite{Wang_2025}. In our setting, a reliable evaluator must (i) give a score to generated ACR bot comment with concrete PR context, and (ii) map that interpretation to a simple usefulness label comparable to the industrial ground truth.

This strategy also follows the shared setup in Section~\ref{sec:shared_eval_setup} for inputs, criteria, and decision mapping. 



\subsection{Evaluation Protocol and Metrics}\label{sec:evaluation-protocol}
We evaluate each automated approach by comparing its binary usefulness prediction against the human-labeled ground truth label
provided in the dataset~\cite{cihan2024automated}.

\textbf{Evaluation Unit.}
The unit of evaluation is a single bot-generated PR comment. Each comment
is independently scored by the evaluator and mapped to a binary
decision using the corresponding binary variants.

\textbf{Primary Metric.}
Following the dataset’s framing, we report agreement ratio:
Agreement = $\#(\widehat{y} = y){/N}$, where $y$ is the human label, $\widehat{y}$ is the automated prediction, and $N$ is the
number of evaluated comments.

\textbf{Aggregation.}
We report results across the overall agreement ratio per model and methodology approach
to capture model variability and evaluation difficulty.

\subsection{Implementation Details}\label{sec:implementation-details}
We implement both automated evaluators as deterministic pipelines that consume the
same comment-level inputs and differ only in rubric definition.

\textbf{LLM Execution Settings.}
All LLM calls are executed with the default temperature and a fixed system prompt. We retained vendor-recommended defaults to reflect realistic industrial deployment conditions, where models are typically used without enforced deterministic decoding \cite{kour2025think}. Because each evaluator was executed once per comment and no repeated sampling was performed, setting the temperature to 0 would have imposed artificial determinism on a single run. Preserving the models’ default stochastic behavior therefore provides a more representative view of practical usage, while robustness under repeated runs and alternative temperature settings remains future work.

\textbf{Token Budget and Truncation.}
To comply with model context window limitations, we truncated PR diffs only when the total input length exceeded the maximum context window size supported by the respective LLM. This affected 227 PRs (approximately 8.7\% of the dataset). In such cases, we preserved the PR content in its original order up to the context window boundary and removed only the remaining tail portion that exceeded this limit.

\textbf{Reproducibility Notes.}
All intermediate artifacts (reasonings, prompts, and raw
model outputs) are logged per comment to support auditability and error analysis.

\section{Results}\label{sec:results}
This section presents the empirical results of the proposed automated evaluation approaches and reports their alignment with human-labeled ground truth. We first present the overall agreement ratio across three models and two methodologies presented in Section \ref{sec:section4}. Further, we analyze the F1 score, precision, recall, and Matthews correlation coefficient (MCC) score for one result with best agreement ratio across all evaluations. We report MCC in addition to the other metrics because the dataset is imbalanced, and MCC provides a more reliable evaluation measure under class imbalance \cite{chicco2020advantages}.


\begin{table}[h]
\centering
\caption{Agreement ratios of automated evaluators against human-labeled ground truth across models and rubric variants.}
\label{tab:agreement_summary}
\begin{tabular}{lcccc}
\toprule
& \multicolumn{2}{c}{\textbf{G-Eval}} & \multicolumn{2}{c}{\textbf{LLM-as-a-Judge}} \\
\cmidrule(lr){2-3} \cmidrule(lr){4-5}
\textbf{Model} & \textbf{Binary} & \textbf{Likert} & \textbf{Binary} & \textbf{Likert} \\
\midrule
GPT-5.2        & 0.45 & \textbf{0.59} & 0.44 & 0.58 \\
GPT-4.1-mini   & 0.48 & \textbf{0.62} & 0.57 & 0.61 \\
Gemini-2.5-pro & 0.51 & \textbf{0.57} & 0.53 & 0.55 \\
\bottomrule
\end{tabular}
\end{table}

\begin{table}[h]
\centering
\caption{Detailed evaluation metrics for the best agreement ratio achieved method (G-Eval - Likert) approach across models.}
\label{tab:llm_judge_metrics}
\resizebox{\linewidth}{!}{%

\begin{tabular}{lccccc}
\toprule
\multicolumn{6}{c}{\textbf{G-Eval (Likert)}} \\
\cmidrule(lr){2-6}
\textbf{Model} & \textbf{Agreement} & \textbf{F1} & \textbf{Precision} & \textbf{Recall} & \textbf{MCC} \\
\midrule
GPT-5.2        & 0.59 & 0.72 & \textbf{0.67} & 0.78 & -0.0020 \\
GPT-4.1-mini   & \textbf{0.62} & \textbf{0.76} & 0.66 & \textbf{0.90} & -0.0590 \\
Gemini-2.5-pro & 0.57 & 0.68 & \textbf{0.67} & 0.69 & \textbf{0.0106} \\
\bottomrule
\end{tabular}
}
\end{table}

\subsection{Performance of G-Eval}\label{sec:results_geval}

We evaluate the agreement between G-Eval and human-labeled annotations using both a direct binary rubric and a Likert-based rubric that is mapped to binary decisions via the thresholding rule described in Section~\ref{sec:Likertscale}. As summarized in Table~\ref{tab:agreement_summary}, the Likert-based variant consistently achieves higher agreement with human labels than the direct binary variant across all evaluated models. For \texttt{GPT-5.2}, agreement increases from 0.45 under the binary rubric to 0.59 with the Likert-based formulation. Similarly, agreement for \texttt{GPT-4.1-mini} increases from 0.48 to 0.62, and for \texttt{Gemini-2.5-pro} from 0.51 to 0.57.

Across models, \texttt{GPT-4.1-mini} achieves the highest agreement under the Likert-based G-Eval configuration (0.62), followed by \texttt{GPT-5.2} (0.59) and \texttt{Gemini-2.5-pro} (0.57). While the absolute differences between models are moderate, these results indicate that the choice of evaluation backend can meaningfully affect agreement outcomes even when the evaluation rubric and input data are held constant.

The consistent improvement observed with the Likert-based configuration suggests that a graded rubric allows the evaluator to express borderline cases more reliably than a single-step binary decision. The subsequent deterministic mapping from Likert scores to binary labels appears to reduce ambiguity relative to forcing an immediate binary judgment, leading to higher alignment with developer-provided annotations. Nevertheless, overall agreement remains moderate, indicating that rubric design and model choice can influence, but do not fully resolve, the challenges of automated usefulness evaluation in industrial code review settings.

To further characterize the behavior of the best-performing configuration, we examine the detailed classification metrics reported in Table~\ref{tab:llm_judge_metrics}. For the Likert-based G-Eval configuration with \texttt{GPT-4.1-mini}, the evaluator achieves an F1 score of 0.76, precision of 0.66, recall of 0.90, and an MCC of $-0.0590$. Across models, precision values are relatively similar, whereas recall varies substantially, ranging from 0.69 for \texttt{Gemini-2.5-pro} to 0.90 for \texttt{GPT-4.1-mini}. The high recall indicates that the judge frequently predicts the \textit{useful} label and captures a large fraction of the \textit{fixed} comments, while the lower precision reflects a higher rate of false positives on \textit{wontFix} comments.

Despite moderate agreement ratios and relatively strong F1 scores, the near-zero (and slightly negative) MCC values indicate weak correlation once class imbalance is taken into account. This discrepancy suggests that agreement and F1 alone may overstate evaluator reliability in this setting, as label distributions and ground-truth confounding where \textit{wontFix} reflects organizational or contextual constraints rather than purely technical non-usefulness limit calibrated discrimination. Overall, these results reinforce that rubric choice and model selection influence G-Eval behavior, but do not fully overcome the inherent challenges of automated usefulness evaluation in industrial code review contexts.

\subsection{Performance of LLM-as-a-Judge}\label{sec:results_llmasajudge}

We evaluate the agreement between the LLM-as-a-Judge approach and human-labeled annotations using both a direct binary rubric and a Likert-based rubric that is mapped to binary decisions via the thresholding rule described in Section~\ref{sec:Likertscale}. As reported in Table~\ref{tab:agreement_summary}, the Likert-based configuration consistently achieves higher agreement than the direct binary formulation across all evaluated models. Under the Likert rubric, \texttt{GPT-4.1-mini} achieves the highest agreement (0.61), followed by \texttt{GPT-5.2} (0.58) and \texttt{Gemini-2.5-pro} (0.55). A similar pattern is observed when comparing binary and Likert variants directly: agreement increases from 0.44 to 0.58 for \texttt{GPT-5.2}, from 0.57 to 0.61 for \texttt{GPT-4.1-mini}, and from 0.53 to 0.55 for \texttt{Gemini-2.5-pro}.

Across configurations, \texttt{GPT-4.1-mini} consistently yields the strongest alignment with developer-provided labels for the LLM-as-a-Judge approach. While the absolute differences between models are moderate, these results indicate that the choice of judge backend affects agreement outcomes even when the evaluation rubric and input evidence are held constant.

\section{Discussion}\label{sec:discussion}

\subsection{Implications for Practitioners}\label{sec:discussion_practitioners}
Evaluating the usefulness of ACR bot comments remains a difficult task \cite{torun2026vision}. Usefulness is subjective and context dependent, and developer outcomes such as \textit{fixed} and \textit{wontFix} can reflect non-technical constraints (e.g., prioritization and timing) in addition to comment quality. In our setting, both automated evaluators achieve only moderate agreement with human labels, indicating that these signals are imperfect and should not be interpreted as an objective ground truth proxy without additional validation. Our results provide several practical insights for teams that deploy ACR bots in real-world software development workflows:

    \textbf{Automated scores should be treated as weak signals, not ground truth.}  
    Given the moderate agreement observed in Section~\ref{sec:results}, practitioners should not blindly trust automated usefulness scores as the sole basis for approving, rejecting, or ranking review bot comments. Instead, these scores are better positioned as heuristic indicators that require periodic sampling and human verification.

    \textbf{Risk of false positives and wasted effort.}  
    Imperfect evaluators can label non-actionable or contextually inappropriate comments as useful. If these scores are surfaced directly to the developers or used to drive automation, they may increase review noise, distract reviewers, and reduce trust in both the bot and the evaluation pipeline.

    \textbf{Use automated evaluation for triage, not decision-making.}  
    A more conservative integration is to use automated scores to prioritize what to inspect (e.g., routing low-score or disputed cases to manual review), rather than to automatically accept or dismiss comments. This can reduce evaluation cost while limiting the harm from misclassifications.

    \textbf{Model and rubric choices matter.}  
    Agreement varies across models and rubric variants (Table~\ref{tab:agreement_summary}), implying that practitioners should not assume evaluator performance is stable across backends or prompt settings. Any deployment should include routine re-evaluation after model upgrades or prompt/rubric changes.

    \textbf{Diagnostics require careful interpretation.}  
    Aggregate evaluation scores (e.g., agreement, F1, precision) may hide recurring weaknesses in specific file types, programming languages, or review scenarios. Therefore, when using evaluator outputs to refine bot prompts or rules, teams should complement these numerical metrics with manual inspection of selected PR examples to better understand common error patterns.

\subsection{Implications for Researchers}\label{sec:discussion_researchers}


    \textbf{Limits of weighting schemes and decision thresholds.}  
    While alternative weighting strategies and decision thresholds can influence evaluation outcomes, our findings suggest that tuning these parameters alone may not resolve misalignment when human labels encode contextual and organizational factors. Future work should therefore study not only how different scoring schemes affect alignment, but also when such alignment is fundamentally constrained by dataset characteristics and task ambiguity.

    \textbf{Human-in-the-loop analysis as a diagnostic tool, not a silver bullet.}  
    Analyzing the reasoning or justification texts produced by LLM-based judges can provide valuable insight into evaluator behavior. However, such analyses should be viewed as diagnostic instruments for identifying systematic biases or failure modes, rather than as mechanisms that guarantee reliable automated judgment. Understanding when and why evaluator reasoning diverges from developer intent remains an open challenge.

    \textbf{Cautious use of fine-tuned decision models.}  
    Fine-tuning decision models on labeled review comments may yield more stable or deterministic outputs. At the same time, this approach risks overfitting to labels that themselves reflect workflow constraints rather than pure comment quality. Future research should examine how fine-tuned evaluators generalize beyond their training context and how sensitive they are to label noise and organizational practices.

    \textbf{Interpreting behavioral signals requires contextual grounding.}  
    Post-merge application or non-application of bot comments can provide additional behavioral signals of usefulness. However, such signals are also shaped by factors such as scope, timing, and developer workload. Research is needed to understand how these implicit signals can be combined with explicit labels without reinforcing existing biases.

    \textbf{Generalization as an empirical question, not an assumption.}  
    Although automated evaluators are often presented as general-purpose tools, our results suggest that their behavior may be tightly coupled to specific repositories, teams, and review conventions. Future studies should treat cross-domain generalization as an empirical question and explicitly report where and why evaluators fail to transfer.

    \textbf{Robustness of evaluation under limited context.}  
    LLM-based judges frequently operate under a limited context, such as partial diffs or missing cross-file dependencies. Understanding how evaluators behave under such constraints, and whether certain judgments are fundamentally unreliable in these settings, remains a key open research problem. Addressing this issue is essential for realistic evaluation in industrial environments.


\subsection{Interview with an Industrial Practitioner}

To contextualize the moderate agreement levels observed between automated evaluators and developer-provided labels, we conducted a semi-structured interview with a director of software engineering at Beko who has direct experience deploying and operating AI-assisted code review tools in production workflows. 
We analyzed the transcript using open coding, an analytic process within grounded theory \cite{stol2016grounded}. The first and second authors independently coded meaningful segments and iteratively grouped related codes into higher-level themes through discussion. 

\paragraph{Theme 1 — Developer labels are operational outcomes, not pure judgments of comment quality.}
The interviewee emphasized that labels such as \textit{fixed} and \textit{wontFix} often reflect delivery constraints rather than purely technical judgments about comment quality. Developers may resolve comments quickly to proceed with deployment, particularly under time pressure. As the interviewee explained: ``Especially on the wontFix side, developers may select it very quickly because the priority is often getting the change into production.'' This observation aligns with our quantitative findings of moderate agreement. If labels encode operational priorities such as deadlines or scope limitations, they cannot be treated as fully objective ground truth for automated evaluation.
\paragraph{Theme 2 — Diff-centered evaluation structurally limits judgment quality.}
The interview repeatedly highlighted the mismatch between what automated systems can observe and what developers consider in practice. While evaluator pipelines typically operate over PR diffs and limited metadata, developers reason holistically about runtime behavior, architectural constraints, and safeguards beyond the visible scope of the PR. This mismatch can produce systematic discrepancies: comments that appear technically valid in isolation may be redundant or already addressed elsewhere, and evaluators may recommend defensive changes for rare edge cases that teams intentionally accept due to broader trade-offs.
\paragraph{Theme 3 — Industrial evaluation is multi-dimensional.}
The interviewee emphasized that evaluation in practice is not reducible to accuracy alone. Practitioners consider risk reduction (e.g., preventing production bugs or security issues), alert fatigue, and integration into the software development lifecycle (SDLC). Over-notification with low-priority suggestions may cause critical issues to be overlooked, while high-severity findings may justify higher operational costs. Thus, usefulness depends on severity, operational impact, and process fit rather than a single aggregate performance score.
\paragraph{Theme 4 — Continuous monitoring is desired, but systematic evaluation is rare without automation.}
The interviewee framed review bots as agents that require ongoing monitoring similar to human contributors, arguing that automated tools should be managed analogously to human performance (agent-resource management). The interviewee expanded, “If we manage human resources, we should also manage agent resources. These tools need to be monitored continuously.” At the same time, systematic evaluation is rarely sustained in practice because manual monitoring is time-consuming and difficult to scale. This tension reinforces our motivation: automated evaluation is attractive as a scalable monitoring mechanism, yet its outputs must be interpreted as context-limited and approximate signals rather than definitive ground truth. The interviewee further noted that such challenges are unlikely to be specific to a single organization, as development culture, business priorities, and acceptable quality thresholds vary across companies.

\subsection{Threats to Validity}\label{sec:discussion_threats}

\textbf{Internal validity.}
Several factors may affect the internal validity of our findings. First, some PR titles in the dataset include embedded issue types or contextual cues. Since our evaluation pipelines extract contextual information programmatically, such metadata may unintentionally bias both G-Eval and LLM-as-a-Judge by providing overly strong or misleading signals. Second, our analysis relies on developer-provided usefulness labels as ground truth. These labels are inherently subjective and may reflect workflow constraints such as time pressure, prioritization, or scope rather than the semantic usefulness of a comment, introducing noise and potential biases into the evaluation~\cite{groundTruthDeficiencies, obaidi2025towards}. Third, the prompt design used to operationalize the evaluation rubrics may influence model outputs, as phrasing, instruction structure, and rubric wording can shape how LLMs interpret usefulness criteria. Finally, LLM-based evaluators may exhibit model-driven biases, for example, by overweighting linguistic clarity, PR titles, or surface-level features of comments, which can skew predictions toward \textit{useful} even when the underlying technical value is limited. 

\textbf{External validity.}
The external validity of this study is constrained by the dataset and evaluation setting. Our analysis is based on multiple industrial repositories from a single organization and focuses on one ACR bot, which limits generalization to other development environments, programming languages, review cultures, or CI/CD practices. In addition, the evaluation is conducted using a specific set of LLM backends; alternative models, versions, or rubric formulations may yield different results. 
Moreover, our setting involves an LLM evaluating outputs generated by another LLM: the ACR bot in the reference dataset is based on GPT-4 Turbo \cite{openai_gpt4_turbo_model_docs_2026}, while our evaluators also rely on modern LLMs. 
This introduces a potential coupling effect, where evaluators closer in model family or training period may better align with the generator’s output style, while others may appear to perform worse due to systematic differences in phrasing or calibration. As a result, performance differences across evaluator backends should be interpreted cautiously and not attributed solely to superior evaluation capability.

\section{Conclusion}\label{sec:conclusion}

This paper examined whether the usefulness of ACR bot comments can be assessed using automated evaluators, using an industrial dataset of 2,604 bot-generated PR comments annotated by software engineers. 
We compared a rubric-based LLM evaluator, which is G-Eval, against the LLM-as-a-Judge pipeline; both approaches are applied over the same binary decision and 0-4 Likert scale evaluation. The agreement ratios, F1 scores, precision, recall, and MCC scores are obtained by agreement with the human-labeled annotations.
Across multiple LLM backends, both evaluation strategies achieved moderate agreement with human labels, with overall agreement ratios ranging approximately from 0.44 to 0.62 in our setting. Across these evaluations, we further analyzed the best strategy among four options: G-Eval (0-4 Likert/binary) and LLM-as-a-Judge (0-4 Likert/binary). The best-performing model achieved 0.76 F1 score, 0.66 precision, 0.90 recall, and -0.00590 MCC score. The results also show that performance varies by model and by the choice of methodology.

These findings suggest that automated usefulness evaluation provides only an approximate and partial signal of human judgment, rather than a direct or reliable replacement.
In particular, wontFix labels may reflect non-technical factors (e.g., scope, prioritization, timing) that are not fully recoverable from PR artifacts or repository snapshots, which introduces an inherent ceiling on agreement-based evaluation. Insights from our interview further support this interpretation, highlighting that developer labeling behavior is frequently shaped by workflow pressures and organizational constraints beyond the observable PR context.
As a result, automated metrics should be interpreted cautiously when used for large-scale monitoring or comparative analysis of ACR systems.

\section*{Acknowledgment} \label{sec:Acknowledgment}
 This work has been supported by the ITEA4 GENIUS project, funded by the national funding authorities of the participating countries: https://itea4.org/project/genius.html
 
\newpage
\bibliographystyle{unsrt}
\bibliography{sample-base}

\end{document}